\documentclass[preprint,journal]{vgtc}                     

\onlineid{0}

\vgtccategory{Research}

\vgtcpapertype{please specify}

\title{Game Engines for Immersive Visualization:\\Using Unreal Engine Beyond Entertainment}

\author{%
  \authororcid{Marcel Krüger}{0000-0002-0085-268X},
  \authororcid{David Gilbert}{0000-0002-3334-5836},
  \authororcid{Torsten W. Kuhlen}{0000-0003-2144-4367},
  \authororcid{Tim Gerrits}{0000-0001-9296-7224}
}

\authorfooter{
  \item
  	All authors are with the Visual Computing Institute at RWTH Aachen University. \\
  	E-mail corresponding author: krueger@vis.rwth-aachen.de
}

\abstract{%
One core aspect of immersive visualization labs is to develop and provide powerful tools and applications that allow for efficient analysis and exploration of scientific data.
As the requirements for such applications are often diverse and complex, the same applies to the development process.
This has led to a myriad of different tools, frameworks, and approaches that grew and developed over time.
The steady advance of commercial off-the-shelf game engines such as Unreal Engine has made them a valuable option for development in immersive visualization labs.
In this work, we share our experience of migrating to Unreal Engine as a primary developing environment for immersive visualization applications.
We share our considerations on requirements, present use cases developed in our lab to communicate advantages and challenges experienced, discuss implications on our research and development environments, and aim to provide guidance for others within our community facing similar challenges.
}

\graphicspath{{figs/}{figures/}{pictures/}{images/}{./}} 

\usepackage{tabu}                      
\usepackage{booktabs}                  
\usepackage{lipsum}                    
\usepackage{mwe}                       
\usepackage[natbibapa]{apacite}
\usepackage{tabularx}
\usepackage{mathptmx}                  

\begin{document}


\maketitle

\section{Introduction}
Developing immersive visualization applications imposes specific demands on the software used to create them.
Choosing a software stack greatly impacts the application's interaction, fidelity, and performance and affects the development process.
Historically, there have been tendencies toward creating custom software to achieve immersive visualizations.
These can broadly be divided into two categories: \textit{bespoke software} specific to the applications and \textit{frameworks} that can be used to build applications.
While bespoke software gives much flexibility, it is often connected to a high (re-)implementation effort.
On the other hand, custom frameworks develop centralized functionality into a common code base, which can then be (re-)used to create immersive visualization applications.
Both, however, face a common challenge today:
Maintenance efforts have increased drastically to meet the growing complexity and demands in various settings.
Coincidentally, game engines --- commercial-off-the-shelf (COTS) solutions --- became more powerful, more open, and easier to use.
This is particularly interesting to immersive visualization labs, as the development of such engines shares several key characteristics with immersive visualization applications.
Both require low-latency computations, interactive rendering, ergonomic interactions, and visual fidelity.
The rise of COTS virtual reality (VR) hardware in the early 2010s especially positively impacted the VR-readiness of COTS game engines.
Besides the similar requirements for the end product, game engines have a large focus on tooling and developer efficiency.
To enable low-friction game development that focuses on gameplay rather than technology, these engines are often highly optimized, bring ready-to-use techniques, and are easy to learn.
Therefore, the obvious question is whether game engines could also be utilized to create immersive visualization applications.

In this work, we present our considerations concerning the use of game engines as a key component of the workflow in immersive visualization labs.
We reflect on historic decisions, development choices, and current workflows within our research lab at RWTH Aachen University, which led from custom software to the use of Unreal Engine (UE) \citep{unrealengine}.
Based on these experiences, we suggest and motivate a list of requirements that we deem important when considering if a game engine is suitable for use in immersive visualization labs.
We then present multiple use cases that highlight challenges within our lab and our solutions and discuss how the engine influenced the development in these particular cases.
Finally, reiterating the introduced requirements, we evaluate and discuss Unreal Engine as a tool for immersive visualization applications and summarize our findings.

\section{Related Works and Background}
Utilizing the advantages of immersive environments to enhance the visualization and analysis of scientific data has become an established option for domain scientists in many applications \citep{rao2020immersive,kuhlen2014quo}.
Early on, it became clear that, besides head-mounted displays (HMDs), specialized output devices like CAVEs and other multi-screen projection systems provide unique benefits~\citep{cruzneira1993,jacobsen1995petroleum, Laha}.
Therefore, it was an important objective to support this wide variety of hardware devices via software.
Consequently, hardware limitations dominated the first research and development questions dedicated to immersive visualization.
As more hardware entered the consumer market, research shifted from technical solutions for hardware towards properties such as efficiency, accessibility, and the overall potential of the software applications.
This trend is also noticeable when looking at the development history of visualization frameworks that have their origin in academia, such as the collaborative visualization and simulation environment (COVISE)~\citep{covise}, DIVERSE~\citep{kelso2002diverse}, Vrui~\citep{kreylos2008environment}, and ViSTA~\citep{van2000vista}.
Current development is less hardware- and engineering-driven and more focused towards general workflows, including teaching and marketing~\citep{zimmermann2011immersive}, digital twins~\citep{dembski2019digital}, and modern rendering approaches~\citep{wang2023immersive}.
The development of ViSTA, our former in-house solution, is described in more detail in the following \textit{From ViSTA to Unreal Engine} section.
Besides applications that use broad general frameworks, many applications are based on either bespoke one-off solutions or frameworks that mainly focus on one domain, e.g., neuroscience~\citep{keiriz2017exploring, marks2017immersive}.
An extensive overview of used solutions is out of the scope of this contribution, and we refer the reader to further literature \citep{9673569, klein2022immersive, kraus2022immersive}.

Commercial off-the-shelf game engines are an alternative whose potential was already highlighted by \citeauthor{friese2008using} in \citeyear{friese2008using}.
In recent years, publications frequently made use of game engines to solve individual domain-driven visualization problems~\citep{kruger2023case,huo2021efficient,marsden2020using}.
Additionally, it can be observed that several labs made Unity~\citep{davis2022cavevr,wischgoll2023center,khadka2023immersive,klassner2023campus} and/or Unreal Engine~\citep{Lugrin2012CaveUDKAV,mayer2023ten, khadka2023immersive} a key component of their workflow.
However, only little work exists that provides a general overview and discussion of the potential and limitations of using game engines.
As more labs consider them as a replacement for custom solutions~\citep{flatken2023immersive}, which is additionally supported by personal exchange with colleagues from the community at multiple conferences, such a discussion could be helpful.

Therefore, in this work, we aim to analyze and discuss game engine requirements, potential, and limitations based on the experience gained within several years of development within our immersive visualization lab.
To provide background on how we obtained that experience, we briefly review our lab's most fundamental development efforts before giving a list of requirements we identified as crucial when considering a development environment for immersive applications.

\subsection{From ViSTA to Unreal Engine}
Our immersive visualization lab's work towards more general application development started with creating the ViSTA toolkit published in 2000 \citep{van2000vista}.
It combined knowledge from the Computing Center of RWTH Aachen University, the Research Center Jülich, the Institute of Technical Computer Science, and the Technical Acoustics at RWTH Aachen University.
As institutes were using VR to solve research problems from different domains on diverse hardware platforms, ViSTA was our first attempt at a flexible general VR framework providing immersive, interactive virtual environments with high visual quality.
Considering the diversity of both hardware (from CAVE systems over powerwall setups to head-mounted displays) and software requirements of the various domain sciences, the framework needed to be as platform-independent and flexible as possible.
Therefore, it was initially built upon the commercial C library \textit{WorldToolKit} (WTK) [no longer available] for practicability, but quickly transitioned to an \textit{OpenSG}~\citep{reiners2002opensg}-based backend with a \textit{Visualization Toolkit} (VTK)~\citep{vtkBook} integration. 
ViSTA was further extended by \textit{ViSTA FlowLib}~\citep{schirski2003vista} to handle Large-scale Computational Fluid Dynamics (CFD) visualizations \citep{gerndt2003large}, and the \textit{Viracocha} framework \citep{gerndt2004viracocha} for parallel processing of large CFD data \citep{gerndt2004vr}.
The focus remained on providing low-level controls to programmers by allowing source-code access and high-level support for engineers to design and implement simulations and visualizations for their virtual environments \citep{assenmacher2008vista}.

With performance and broad functionality mainly covered, more advanced analysis interactions were needed, such that the focus shifted towards more sophisticated 3D UI interaction techniques \citep{gebhardt2013extended, gebhardt2016vista}.
Uses thereof were shown in the visualization of room acoustical simulation data \citep{freitag2011visualizing}, probabilistic fiber tracts \citep{rick2011visualization} and air traffic noise \citep{pick2013virtual}.
Due to these extensions and alterations, ViSTA was used in many immersive visualization applications, e.g., in the medical domain \citep{nowke2013visnest, Hanel2014b, knott2014data} as well as general research on immersive visualization \citep{hanel2016visual,Hanel2014a, pick2016design, freitag2016automatic}.

While ViSTA aimed to provide a toolkit with high flexibility, good third-party support, and community development by being open source, specific weak points became apparent early on.
The development process was time-consuming and complex, while user accessibility was low due to an exclusive C++ interface, as noted by \citet{assenmacher2008vista}. 
With the increasing number of additional extensions and rewrites, maintenance of the toolkit source and documentation became even more difficult.
This worsened the already impaired accessibility, leading to high training periods and increased frustration for students and new staff.
At the same time, the need for higher visual fidelity rose as people became familiar with modern rendering capabilities from mainstream media.
Due to the proliferation of consumer VR hardware and its support in COTS engines like Unity and Unreal Engine, individual research had already begun to transition away from ViSTA~\citep{freitag2016automatic} to game engines~\citep{freitag2017approximating, freitag2018interactive}.

After a short development-phase of an in-house successor to ViSTA, the decision was made to reevaluate the use of game engines.
The primary motivations were our initial breakthroughs for supporting our CAVE environment and general indications that COTS game engines had become more accessible to the research community, e.g., \citet{donalek2014immersive,cordeil2016immersive,sicat2018dxr,capece2018graphvr}.
It was clear that the advantages of ViSTA, namely flexibility, performance, extensibility, and adaptability, should be maintained.
At the same time, a potential switch to a COTS game engine allowed us to further evaluate them on accessibility, documentation, community support, and feature set.
The eight properties above were developed into a list of six final requirements, ultimately leading to Unreal Engine's adoption as a base for future development.
As these are strongly connected to our lab, a short description of our work environment and hardware is provided below.

\subsection{The Immersive Visualization Lab at RWTH Aachen University}
The Visualization lab at RWTH Aachen University is a research lab in an academic setting with a dual purpose. 
The academic aspect of lab usage is given through research by the Visual Computing Institute. 
Besides academic research, the lab provides services for other institutes and external cooperation partners, such as industry partners, through the central resources offered by the IT Center at RWTH Aachen University. 
Research is done by academic staff and supported by around ten student workers, while non-academic staff members predominantly provide service requests. 
At any given time, there are around six non-academic staff members and 12 academic staff members, while approximately 24 students start their theses each year.
However, there is a high fluctuation in both categories of staff due to fixed-term employment contracts ranging between three months to six years.

Two categories of immersive visualization hardware are actively operated in our lab.
First, many different VR and Augmented Reality (AR) COTS HMDs from various manufacturers are provided with regular new purchases. 
The HMDs are used in standard tethered or wireless configurations driven by stationary workstations or standalone modes.\\
The second category describes room-mounted displays, which we operate in three different scenarios:
\begin{figure}[ht!]
    \centering

    \includegraphics[width=\columnwidth]{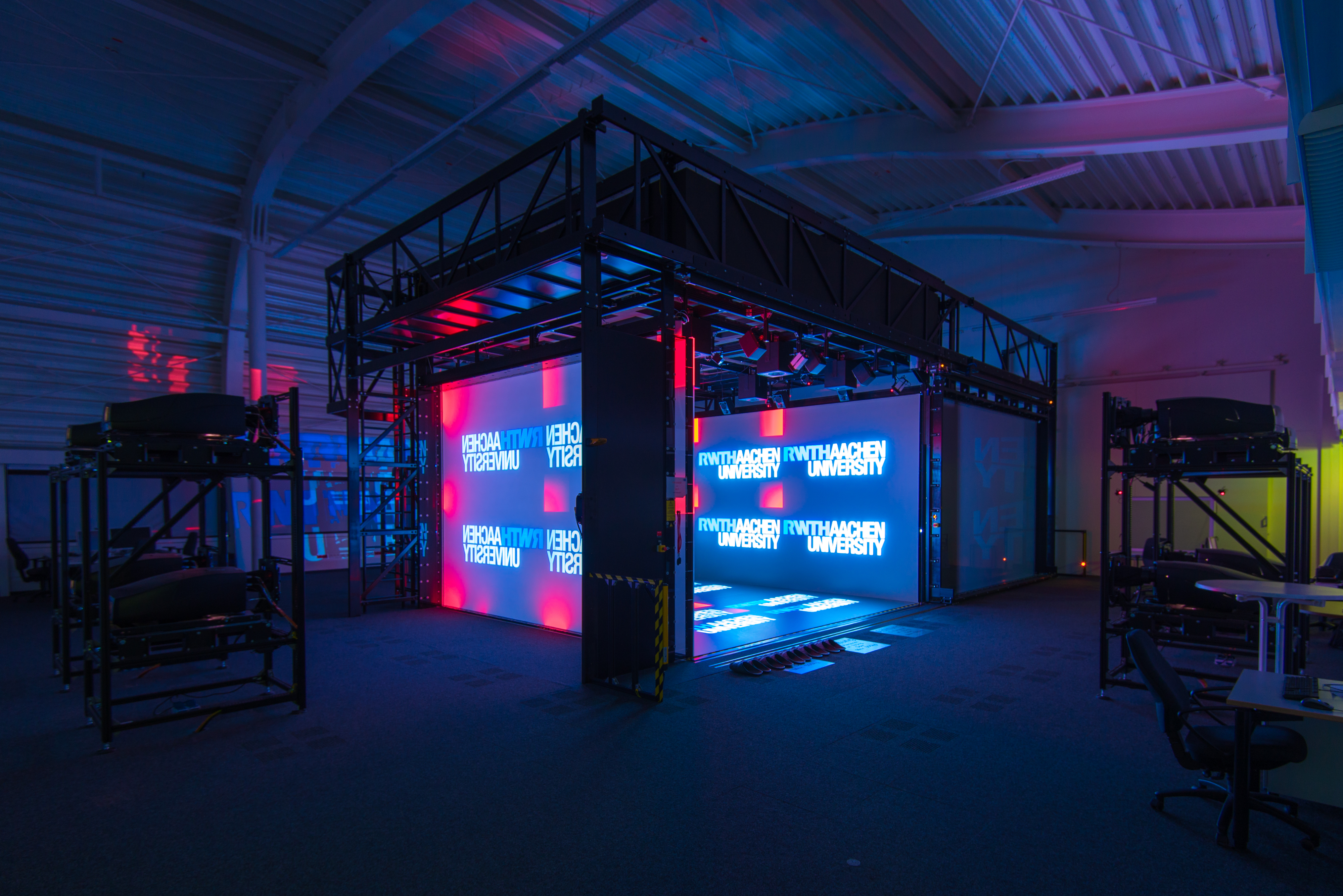}

    \caption{The five-sided CAVE at RWTH Aachen University. The measurements are $5.25$ m x $5.25$ m x $3$ m. Each side is rear-projected by four projectors, with an additional eight projectors under the floor. A 25-node Linux cluster drives projectors.}
    \label{fig:aix-cave}
\end{figure}
The \textit{AixCAVE} is our five-sided $5.25$ m x $5.25$ m x $3.30$ m CAVE system (c.f. Figure \ref{fig:aix-cave}). 
Four projectors drive each side wall, while eight drive the floor due to the bigger area.
A resolution of close to 4K for the side walls and close to $3840$x$4800$ for the floor is achieved (some resolution is lost due to blending).
The AixCAVE is driven by a 25-node Linux cluster, where each node is equipped with two GPUs dedicated to one stereo channel per projector.
On the user side, stereo is realized via active stereo shutter glasses, which are tracked via an optical tracking system from \citet{art}.
The 25th node drives our main workstation, which controls the CAVE operation, synchronization, and input.

We further run a \textit{tiled display wall} consisting of $2$x$3$ displays driven by a seven-node Linux Cluster, providing our users with head-tracked visualizations.
Finally, a \textit{mobile stereo projection powerwall} with head- and controller-tracking is used as a portable immersive visualization solution.
It provides a resolution of $2650$x$1600$ pixels, projects onto a $3.00$ m by $1.70$ m meter screen, and is operated by a standard workstation.
Tracking is realized via optical tracking markers on the shutter glasses and input via a tracked joystick.
\section{Requirements}

Through careful consideration of our lab's specific environment and our extensive hands-on experience in creating immersive visualization applications and frameworks over the years, we have identified six fundamental requirements that software solutions should satisfy to serve as a robust foundation for development.
These cover aspects not only from a strict feature and development perspective but also include elements considering the workflows in our lab, e.g., the frequency of people joining and leaving, their average stay at the lab, and their technical proficiency. 
This results in a mixture of both hard and soft requirements:
\begin{enumerate}
    \item \textbf{Wide Adoption}
    \item \textbf{Large Feature Set}
    \item \textbf{Performance}
    \item \textbf{Accessibility}
    \item \textbf{Extensibility/Adaptability}
    \item \textbf{Flexibility}
\end{enumerate}
In the following, we discuss and motivate each requirement in the context of our lab.
We evaluate how COTS game engines meet these requirements and compare them to the alternative of developing custom in-house frameworks.
While being developed initially in the context of our lab's needs, we did not consider features provided by ViSTA as direct requirements for a game engine.
Instead, we deliberately formulated properties that are important for using game engines as a base for further development.
The fulfillment of these allows us to modify the engine to meet domain- and lab-specific needs while benefiting from an engine's general properties.
Due to this, we believe that they are not only general enough to be applicable for other labs in their decision-making processes but also useful for general application development.
Therefore, we used the same requirements in our case study on using a COTS game engine to develop task-specific applications~\citep{kruger2023case}.
Nonetheless, we encourage reevaluating and modifying the presented requirements by adding, removing, and weighting them to fit individual boundary conditions.

\begin{description}[wide, labelwidth=!, labelindent=0pt]
    \item \textbf{Requirement 1 - Wide Adoption}
    As described in the background section, our lab experiences a nearly full replacement of staff members every six years while students join and leave more frequently.
    People joining shall be productive quickly, and knowledge loss must be minimized when people leave. 
    A wide adoption increases the chance that newcomers already know the engine and code base, allowing them to implement new features and tools right away.
    Compared to custom framework solutions, new work staff or students are often familiar with game engines due to prior work experience or hobbyist projects.\\
    Using a widely adopted engine also increases the chance that cooperation partners use the same engine.
    This leads to faster and easier collaboration as it is clear from the beginning that the project will run on their respective local infrastructure.
    There is no need to install, maintain, and test several software stacks simultaneously to provide interoperability in different locations.
    Widespread use in academic research facilitates peer reviewing and reproducibility.
    Additionally, it allows the integration of new algorithms into one's applications without reimplementation in case a paper's source code is released.
    Finally, a wide adoption entails a large community of active users, which increases the chance for long-term support due to demand.
    Engines often allow the creation and import of functionality as plugins, which can be used to include new functionality into the engine.
    A big and active community facilitates discourse, support, and exchange of experiences and an ecosystem of community plugins.
    This allows labs to benefit from and contribute back to a striving community.

    \item \textbf{Requirement 2 - Large Feature Set} A large feature set is beneficial to speed up the development as existing engine functionality can be used in the application without new implementations.
    Use cases brought into the lab can differ vastly based on research domains: While one project needs photo-realistic lighting with acoustic rendering, the next project requires efficient rendering of flow data and analysis features.
    By utilizing features already available in the engine, spending resources on potential one-off implementations can be avoided.
    Additionally, features are mostly highly optimized and battle-tested solutions the engine vendor provides.
    Future improvements and optimizations are often developed with continuity in mind, such that they seamlessly integrate into existing projects or require only little adaptations.
    Incorporating new features into custom software is often time-consuming enough that additional resources for testing and optimizing are hard to justify, especially if there is a limited need in the foreseeable future.
    Thus, using COTS engines can significantly increase the quality compared to custom solutions if no extensive testing procedures exist.
    Besides functionality that can be used in an application, an important part of an engine's large feature set is development tools.
    Tooling in COTS game engines spans various areas, from scene authoring, performance analysis, or orchestration to development efficiency and test suites.
    It is often considered a secondary aspect that frequently falls short when developing custom solutions as it takes up large amounts of resources without any immediate return, thus making it hard to justify spending a large number of development resources on it.
    Good tooling, however, can drastically speed up implementation, optimization, building, and debugging.
    Considering tooling as part of the large feature set that COTS game engines provide is extremely important and should not be underestimated.
    \item \textbf{Requirement 3 - Performance}
    Performance is critical when developing immersive visualization solutions, as those often deal with large data sets or complex data types.
    When choosing an engine to use as the primary engine for a lab, the engine must handle various domain-specific applications with vastly different performance demands.
    As demands of new hardware and techniques change drastically, we require an engine that is mature and optimized for general performance regarding rendering and interactions on immersive devices.
    Today's hardware market shows rapid growth in hardware specs, making performant rendering more challenging.
    Accordingly, the complexity of rendering techniques increases steadily through new algorithms and hardware solutions.
    Game engines are often highly optimized and offer implementations that use hardware accelerations and modern techniques implemented in cooperation with hardware manufacturers.
    For custom-developed solutions, it is often difficult to achieve the same optimizations due to the limited resources available compared to engine vendors.
    This is especially true when considering mobile platforms that additionally require implementations to be energy efficient.
    \item \textbf{Requirement 4 - Accessibility}
    Good accessibility to the engine allows new workforce to be productive quickly.
    We define accessibility as two significant properties which are not mutually exclusive.
    On the one hand, we desire an accessible engine with regard to the availability of resources, such as documentation and tutorials.
    On the other hand, we also define accessibility as the amount of knowledge and skills needed to use an engine productively.
    An accessible engine allows users of different knowledge levels to develop applications on their own.
    Especially, features such as easy-to-use scripting languages, visual programming capabilities, and useful default configurations can help users to get into an unfamiliar engine.
    An accessible engine also allows collaboration partners to contribute to the development of applications and use the software in their facilities.
    Domain scientists can already try out the first steps of an application on their own and thus can see the benefits hands-on.
    As shown in our previous case study \citep{kruger2023case}, modern game engines allow even novel users to get first results.
    However, in later stages, they often need and benefit from the expertise of visualization experts.
    By choosing an accessible engine, labs can support domain scientists who already made their first steps independently.
    Compared to custom software, most game engine vendors recognize accessibility as a key ingredient in increasing their market share.
    Therefore, major engine vendors spend many resources on developing features, tools, and learning resources to onboard new developers and welcome them into their ecosystems.
    In the context of custom software for scientific use, this is often a low priority due to reduced resources or perceived low importance.
    \item \textbf{Requirement 5 - Extensibility/Adaptability}
    Since immersive visualization labs often operate special hardware not found on the consumer market, we require an engine that allows extension and/or adaption to support such systems.
    If extension and/or adaption are necessary, the required work should be as minimal and non-invasive as possible.
    Otherwise, the maintenance burden, whenever updates to the engine are released, is significant, which can lead to less agile updates.
    Besides the support of special hardware, it enables the integration of algorithms deep into the engine.
    This is especially relevant in cases where maximum performance is required, such as algorithms that must scale, are sensitive to timings, or are relevant for the function of the engine as a whole.
    The ability to extend the engine at a low level allows the implementation of these algorithms in-house and the embedding of existing libraries and algorithms into the engine.
    Compared to custom engines developed for a specific purpose, however, it must be noted that implementing these modifications into game engines is often more complex.
    The enormous size of modern engines can make it challenging to integrate new or modify existing features.
    
    \item \textbf{Requirement 6 - Flexibility}
    With the rise of HMDs in the consumer market, the number of potential platforms increases yearly. 
    Many display and interaction devices now run on mobile processors and operating systems such as Android, further increasing the differences between the platforms. 
    A suitable game engine for use in a lab should be able to run developed applications on various platforms without needing modifications.
    It allows the deployment of applications on the hardware of cooperation partners without restrictions on compatible hardware. 
    Giving cooperation partners the freedom to make the best decision based on required fidelity, features, and budget.
    Being designed for consumers and often entertainment, hardware manufacturers and game engine companies have a genuine interest in supporting their respective products.
    Therefore, plugins and extensions often already exist at launch for new hardware, which can be easily used in cutting-edge research. 
    Due to this, there are only two viable options for custom software: New platforms cannot be supported, or the development and maintenance effort increases significantly.
    The diversity and heterogeneity of the hardware market pose a considerable challenge for custom software.
    Enormous resources must be spent to maintain all the different platforms, hardware architectures, and SDKs needed to operate such hardware.
    This limitation makes it harder to support new devices and replace broken devices and imposes hardware restrictions on cooperation partners that want to use the applications in the field or their facilities.
\end{description}
\section{Unreal Engine in the Lab}
In the previous section, we presented requirements we deem important for using COTS game engines as the primary development tool for an immersive visualization lab. 
UE was already widely adopted in the gaming industry, and increased usage in virtual production scenarios brought initial support to clustered rendering to the engine core. 
This extended the engine's flexibility enough to yield initial breakthroughs in compatibility with our CAVE environment. 
Therefore, a thorough consideration of UE as a primary lab engine was undertaken.
The provided feature set was unquestionably large enough and constantly growing, while performance remained a major UE focus.
With the addition of the Blueprint scripting system in UE 4, the engine became accessible even for new developers. 
At the same time, the extensibility and adaptability still allowed the development of complex, specialized techniques.
In the following section, we explore various aspects of utilizing UE in our lab and describe our experience in how their development relates to our defined requirements.
We divided them into three subsections.

The first subsection, \textit{Using Unreal Engine}, highlights our experiences gained from employing UE to develop applications that required data visualization. We utilized the engine's native features or available plugins to realize these applications. We focus on the following examples:
\begin{enumerate}
    \item Previous Case Study Results: Findings from a previous case study about COTS hard- and software in immersive visualization.
    \item Use Cases with Offline Pre-Processing: Two use cases employed different offline pre-processing levels to create immersive visualizations.
    \item Dynamic Data Loading Use Case: Facilitating the dynamic loading of data without offline pre-processing.
\end{enumerate}
The second subsection, \textit{Extending Unreal Engine Functionality}, explores the experiences we gained by extending the functionality provided by UE. It covers the following topics:
\begin{enumerate}
    \item Adaptation for AixCAVE: How we extended and adapted UE to support it on the AixCAVE.
    \item Large Scale Line Rendering Use Case: A use case that compelled us to enhance the engine's functionality to achieve large-scale line rendering in our applications.
    \item Integration with NVIDIA OptiX: A non-trivial integration between the NVIDIA OptiX library \citep{parker2010optix} and UE enabled us to perform partial OptiX accelerated rendering in UE scenes.
\end{enumerate}
The third subsection, \textit{Impact on Daily Work}, gives an overview of how switching to UE impacted our day-to-day workflows in the immersive lab. It covers the following topics:
\begin{enumerate}
    \item General Tooling: How tools included in UE benefited our daily work.
    \item Teaching Activities: The impact UE's wide adoption has on theses and lab courses, our main teaching activities with UE.
    \item Immersive Visualization Services: How UE improved many of our visualization services offered to third parties and other institutes.
\end{enumerate}

\subsection{Using Unreal Engine}
The large feature set, good performance, and flexibility allowed us to use the engine for several projects without any modifications.
UE provided us with excellent support through a wide range of hardware. 
Students, researchers and service personnel use a mixture of SteamVR-based headsets like the Valve Index, Vive Pro 2, and the Windows Mixed Reality-based HP Reverb G2s (+ Omnicept Edition).
Extra HP Reverb G2s are available to hand out, as those are relatively easy to plug and play without tracking systems set up due to their inside-out tracking. 
In addition, standalone headsets like the Pico Neo 3 Pro Eye, Pico 4, Meta Quest, and Vive Focus are used to research non-tethered approaches and streaming techniques. 
While these devices are mainly Android-based, they all use different backends and often require unique plugins. 
Lastly, some research projects require AR-capable devices like the Hololens 2.
UE supports the diversity of different HMDs with various software backends and operating systems, as many official manufacturer plugins for UE exist. 
UE is flexible enough to target Linux, Windows, iOS, and Android devices per default, requiring no engine or application modifications for deployment.
With comprehensive support for the OpenXR standard by \citet{openxr}, UE also adheres to open standards without relying on different manufacturer plugins. 
The native flexibility UE provides allows us to develop for all mentioned devices simultaneously without any additional work.

\begin{figure*}[ht!]
    \centering

    \includegraphics[width=\linewidth]{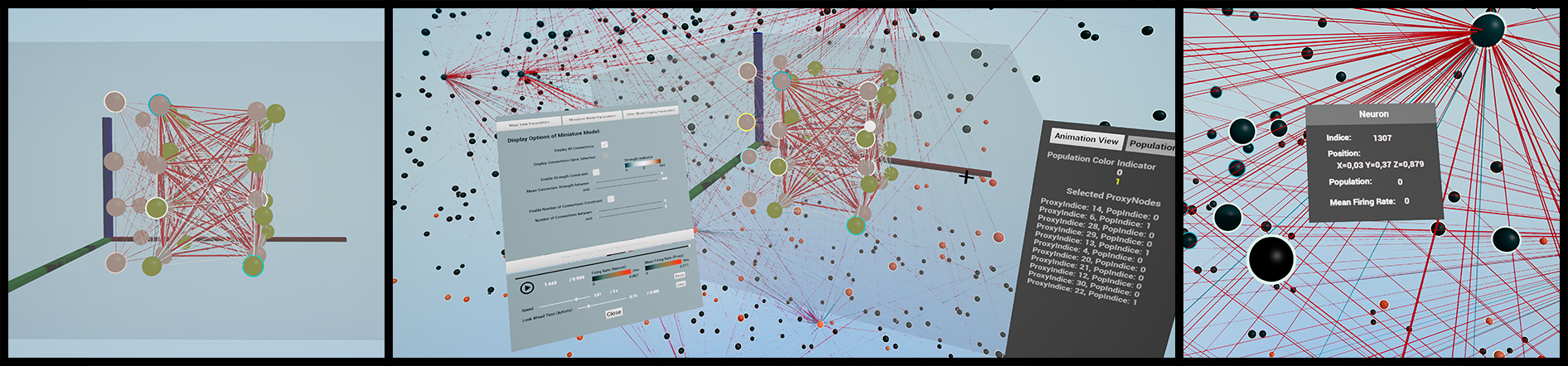}
    \caption{Left: Miniature view of the binned network data. Clicking on a node enables the rendering of the whole area in the main view. Middle: Main view, miniature view, and system control elements. Right: Main view with info UI. Image provided by \citet{kruger2023case}}
    \label{fig:cots_neuronal}
\end{figure*}

\paragraph{1. A Case Study on Providing Immersive Visualization for Neuronal Network Data Using COTS Soft- and Hardware}
As seen in the previous paragraph, game engines make developing for virtual reality hardware comfortable when sticking to commercial off-the-shelf hardware.
We, therefore, investigated how easy it is to use UE to also develop the needed software without specific immersive visualization knowledge \citep{kruger2023case}. 
The key question was if creating an immersive visualization application could be possible while only relying on na\"ive methods.
The motivating initial and provoking thought was that development has become so easy that immersive visualization developers are no longer needed, and domain scientists can develop such applications independently.
In the paper, the goal was to use UE to develop an application that allowed to explore data from a neuronal simulation. 
Activity and topological data were visualized via an UE application in an HMD.
The two underlying concrete research questions were a) can domain experts create such applications on their own without advanced programming knowledge, and b) can immersive visualization labs benefit from using a game engine to create such applications?
The developed application used a linked-view approach based on Shneiderman's mantra of overview first, details on demand.
The network was visualized in a world-in-miniature that represented the topology and activity of the network in a spatially binned representation.
The user could then toggle specific regions and see the detailed topology and activity in the main view (c.f. Figure \ref{fig:cots_neuronal}).
As described in the paper, we deliberately stuck to na\"ive implementations that could be achieved by following online tutorials to investigate how far these techniques can get a developer.
While a running prototype was created very quickly, the resulting application clearly showed severe performance issues that rendered the application unusable for medium-sized data sets, mainly caused by Blueprint implementations and the overuse of na\"ive data structures.
In general, the findings of this case study aligned with the broad experiences gathered by using UE as the primary game engine for our lab:
While the benefits outweigh the drawbacks, UE is no silver bullet, and most drawbacks can be compensated by using good software development practices and methods described in the other use cases.
Overall, it has shown us that UE can help immensely when developing applications, but a lot more is needed until domain scientists can use it on their own.
It reinforced our hopes that using a COTS game engine makes development easier but also gave us indications of which areas and use cases we have to take particular care of, especially when giving tasks and projects to developers that are not experienced yet.

For our next use cases, which cover a common topic in immersive visualization, we were able to use UE out of the box:
The rendering of already generated iso-surfaces.

\paragraph{2. Use Cases with Offline Pre-processing}

\begin{figure*}[ht!]
\centering
\includegraphics[width=0.448\linewidth]{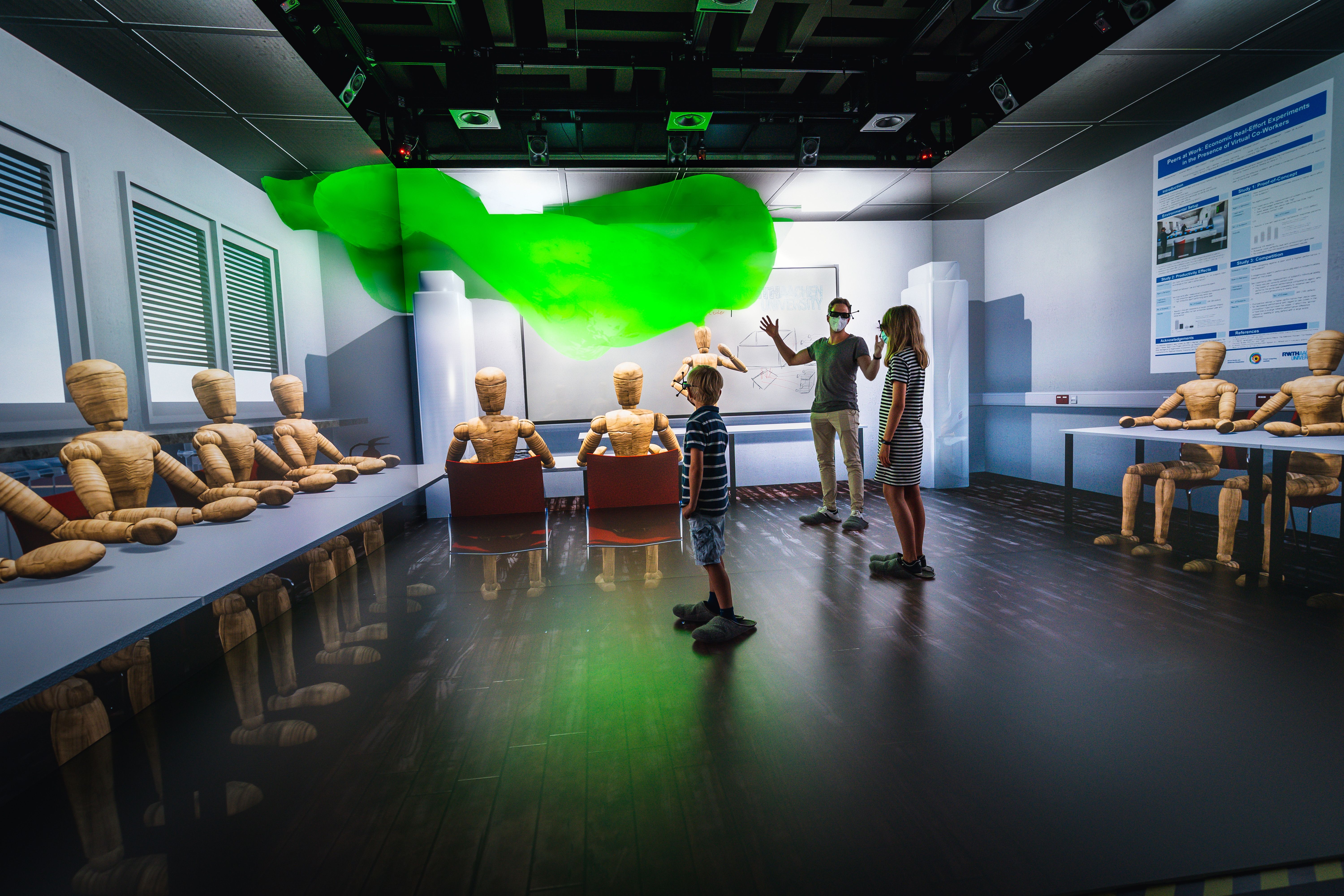}
\hfill
\includegraphics[width=0.532\linewidth]{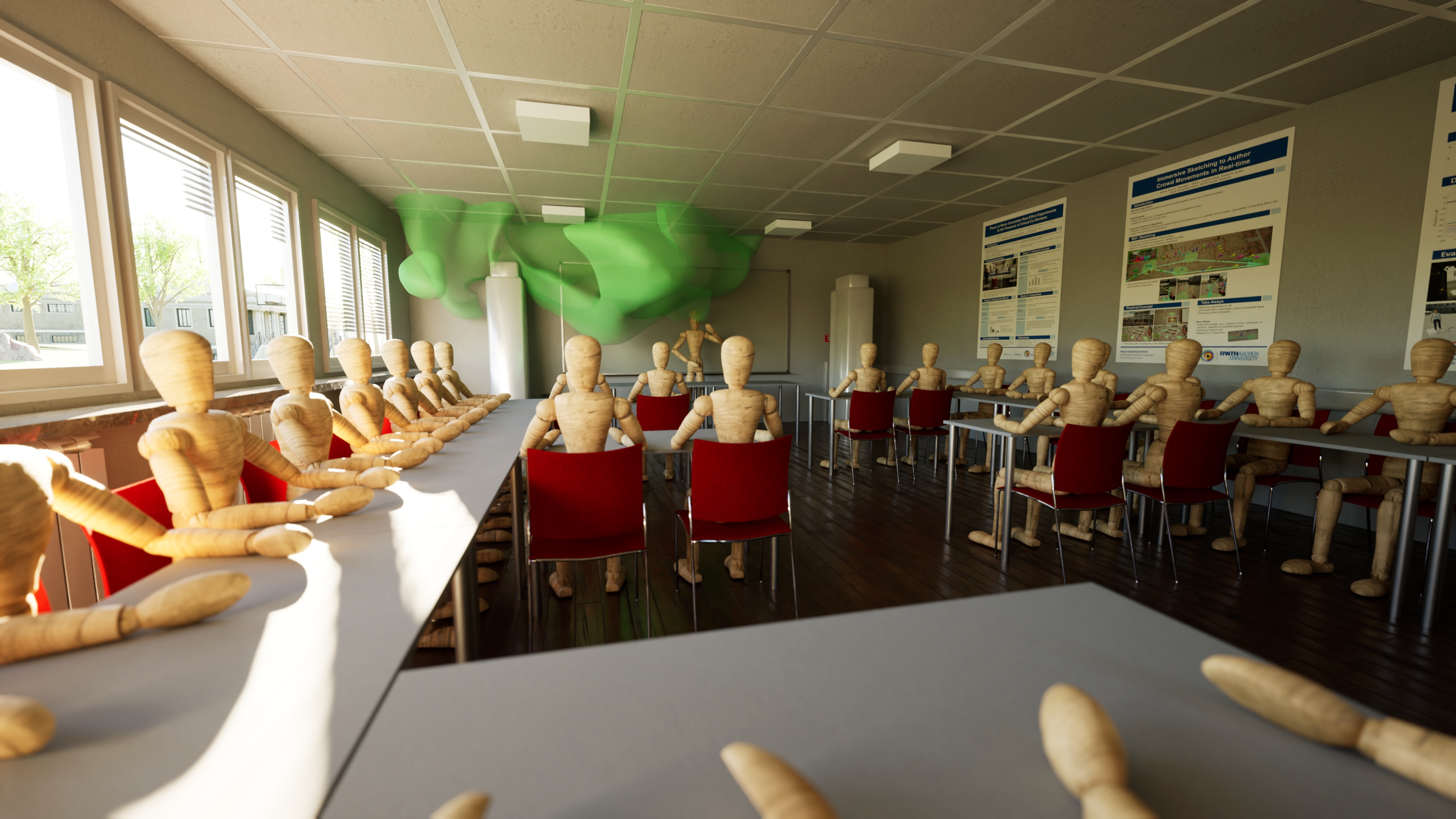}
\caption{Left: The aerosol visualization is shown in our CAVE to explain the mechanics of aerosol spreading to school kids. Right: A pathtraced image of the aerosol visualization from a student's perspective, generated via UE's offline pathtracing renderer.}
\label{fig:aerosolcave}
\end{figure*}

At the beginning of the SARS-CoV-2 pandemic, much research was conducted to aid understanding of how the disease behaves.
One key element was to gain insights into the mechanics of virus spread.
While researchers figured out that the main transmission paths were airborne due to aerosols emitted into the air, getting an intuitive feeling of how aerosols spread is difficult due to the invisible nature of these particles.
In collaboration with EON Research Center (ERC) at RWTH Aachen University, an application was developed to make these mechanisms visible.
The ERC simulated the spread of aerosols in a typical German classroom via fluid simulation.
Two tasks were necessary to create an immersive experience that let users observe the aerosol spread.
First, a visually pleasing representation of the classroom was needed.
Second, the simulation results had to be integrated into the UE application.
To tackle the first task, we exported the boundary regions as a single untextured mesh that was used to align textured meshes with higher fidelity in the scene.
To visualize the simulation results, we opted for an iso-surface representation of the aerosol concentrations in the air.
Due to the offered data formats, a direct import from ANSYS CFX into UE was impossible.
A time-efficient approach was chosen, and we first exported the simulation data from ANSYS CFX as ensight files to disk.
The data was then loaded into the open-source visualization software \textit{Paraview} \citep{ahrens200536}, where an automation script extracted all iso-surfaces into a gltf file for each timestep.
It was then imported into UE at runtime via a community plugin and animated by an interactive flipbook animation. 
While the pre-processing was done in an offline step, the loading and rendering of the aerosol clouds were done in real-time.
Figure \ref{fig:aerosolcave} left shows the application in our CAVE with two school children and one visualization expert explaining the application in the immersive environment. 
Using UE not only allowed us to get from concept to the first interactive version in two weeks but also enabled us to render high-quality pathtraced images.
Figure \ref{fig:aerosolcave} right shows the same scene from a student's perspective rendered with UE's high-performant offline pathtracer without any changes.
While the discussed solution does not allow the user to dynamically change the iso-value at runtime, we chose this approach due to its simplicity and limited implementation time.
While it is possible to also import the raw data into UE and perform runtime triangulation, UE gave us the flexibility to use native features to find a fast workflow that was sufficient for the use case.
The resulting application was used for a multitude of demos, and the pathtraced renderings were used as images in social and print media by external parties.
Especially the short amount of development time allowed us to produce convincing results quickly while the topic was important and still extremely relevant.
The short development time was predominately achieved using advanced scene authoring features to build a visually pleasing classroom scene from simple boundary data and UE's large feature set and community plugins.
By sticking to good practices, such as implementing computation and data-heavy operations in C++, object pre-allocation, object reuse, and a simple architecture, we achieved good results without encountering performance issues, as opposed to the limitations in the na\"ive implementation of the neuronal simulation visualization.

\begin{figure*}[ht]
    \centering
    \includegraphics[width=\linewidth]{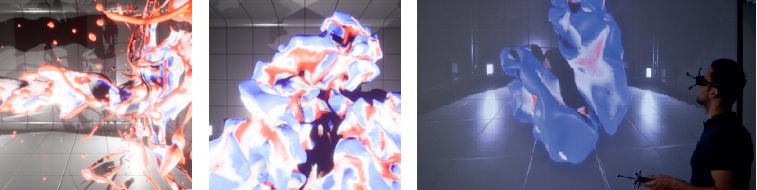}
    \caption{Left/Middle: Application screenshots of the iso-surfaces in the virtual lab. The user can change the lights' positions and move through time to observe the temporal behavior. Right: Scientists observing the combustion dynamics using our mobile powerwall.}
    \label{fig:flame_vis}
\end{figure*}

A similar project was the visualization of combustion processes developed to support research at the Institute for Technical Combustion at RWTH Aachen University.
It allowed domain scientists to investigate the behavior of early flame kernels in an immersive environment to better understand the transition from laminar to turbulent flow as shown in Figure \ref{fig:flame_vis}.
Users could interact with the time-varying data in a virtual lab setting.
Again, the data was represented as fixed iso-value iso-surfaces, replayed through time using flipbook animations.
However, a slightly different workflow was used compared to the aerosol visualization.
While the pre-processing was similarly done using Paraview producing iso-surfaces as gltf files, these were imported into UE via the asset importer, which converts the data offline into an UE uasset file.
Compared to run-time loading, the advantage of this approach is that UE converts the data into its native format and applies some processing allowing for features such as mesh reduction, normal recomputation, and offline generation of signed-distance fields needed for, e.g., light baking.
Therefore, all features and optimizations of UE can be applied to the mesh, and the data is integrated into packaged executables.
The use of native static UE meshes had a positive impact on the performance of the application due to the use of the optimized static mesh rendering pipeline.
Advanced features like mesh reduction and light baking significantly reduced the run-time complexity of the rendering process.
While this approach requires all data to be available beforehand compared to the run-time loading of mesh data, it results in an extremely easy-to-deploy self-contained package that runs on many different systems, from desktop PCs to different VR headsets.
This is extremely helpful when one of the goals is deploying the software at other facilities besides our lab.

\begin{figure}[ht!]
    \centering

    \includegraphics[width=\columnwidth]{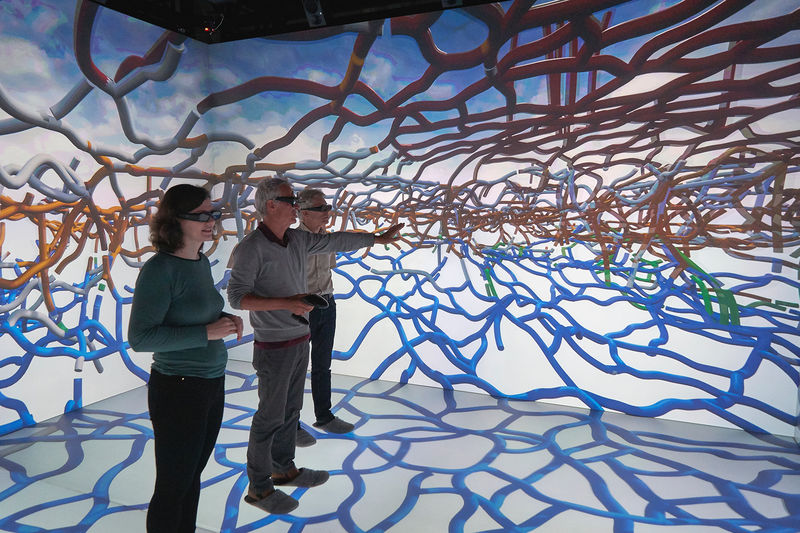}
    \caption{Domain experts discussing a cytoskeleton in the AixCAVE. The filament parts' coloring can be interactively modified with the flystick.
    }
    \label{fig:cytoskeleton_users}
\end{figure}

\paragraph{3. Use Cases with Dynamic Processing}
As is common in visual analytics applications, exploration may require dynamic generation of interesting structures such as iso-surfaces instead of relying on pre-processing of the data.
The focus of the cytoskeleton project was to provide experts from the Institute of Molecular and Cellular Anatomy of RWTH with an application supporting the analysis and human-guided classification of filament structures of cytoskeletons \citep{windoffer2022quantitative}.
To give flexibility to the domain scientists, an immersive visualization application for both CAVE and HMD was required.
Therefore, experts were free to choose between the CAVE for, e.g., collaborative analysis (c.f. Figure \ref{fig:cytoskeleton_users}) or use HMDs at their facilities.
Contrary to the other approaches presented so far, the application did not rely on using pre-generated meshes.
Instead, a runtime mesh generation approach was used to simplify the usage of the software, i.e., to have a self-contained solution for the domain scientists that does not require pre-processing via third-party software.
The cytoskeleton visualization was provided as a csv file describing the network structure of the cytoskeleton.
The mesh was then dynamically generated with UE's \textit{ProceduralMesh} generator.
While using the application, the domain scientist could manually classify certain segments for further use by re-exporting the labeled data via csv files.
The ability to generate and change meshes via native engine functionality in UE was essential.
It allows to load unsupported mesh file formats at runtime by implementing a parser that translates the mesh encoding into procedural meshes.
Thus, it is another example of how UE can be used to quickly develop immersive visualization applications running on COTS hardware, which was possible due to the high accessibility, performance, and large feature set.
%

\subsection{Extending Unreal Engine Functionality}
While the previous section gave an overview of applications we could develop by simply using UE's native functionality of available plugins, the following section describes three cases that required the extension of the engine either through modifications of the engine itself or through plugin development to provide immersive visualization capabilities.
\paragraph{1. Adaption for AixCAVE}
As mentioned before, the first challenge when we started using UE was supporting all the hardware our lab operates.
The support of our CAVE system was an essential requirement.
At the time of the transition to UE, the engine started to support multi-display, multi-node renderings by providing nDisplay \citep{nDisplay} as a core plugin maintained by Epic Games.
As nDisplay was originally developed to power virtual productions, the overlap with clustered rendering in a CAVE environment was large enough to provide a good base for us.
However, our CAVE environment is non-standard in two main aspects.
First, using a Linux cluster is not standard compared to most users that use UE with nDisplay for gaming or virtual productions.
Second, our AixCAVE runs a tracking system via \citet{art}, which had to be integrated into nDisplay to provide correct rendering perspectives to the user.
Due to the access-to-source policy by Epic Games, we could adapt the existing nDisplay implementation to our needs.
This adaption led to a contribution to an early working Linux version of the nDisplay plugin based on OpenGL rendering while using source-built Linux binaries of \textit{VRPN} \citep{taylor2001vrpn} for tracking.
The extensibility of UE's plugin system allowed us to integrate those third-party binaries, yielding a working version for our CAVE up to and including UE 4.23.
However, UE deprecated OpenGL support in favor of Vulkan in version 4.24. 
While this is a welcome change, it was challenging because nDisplay initially did not support Vulkan as a rendering backend.
Support of the newest versions was always an important aspect, both to be compatible with community developments and to benefit from new features.
To remedy the issue, we developed Vulkan support for the nDisplay plugin. 
While challenging initially, as the previous implementations relied on hardware barriers \citep{nvapi} for frame- and genlocking, we solved the issue by falling back to software-based TCP barriers to keep projectors synchronized. 
To our knowledge, there is still no straightforward way to access the same barriers with Vulkan.
However, coupled with frame-locked GPUs, Unreal's na\"ive Vulkan Vsync implementation, and Vulkan's FIFO presentation mode, this yielded a tearing-free stereo projection on our system.

After developing the first working version of nDisplay's Vulkan support, we provided the implementation as a pull request to the community and Epic Games.
Vulkan support for nDisplay was finally added officially in the 4.27 release of UE, thus allowing us to use the official release of UE instead of our fork.
The extensibility and adaptability of UE allowed us to develop a custom fork of the engine that ran on our specialized hardware even without official support.
The later addition of the \textit{LiveLink} UE plugin made it possible to avoid using VRPN in favor of a direct coupling to ART, our used tracking system, by adapting the existing plugin.
This allowed us to run UE as the primary engine of our lab up to the latest UE 5 version.

\paragraph{2. Efficient Line Rendering}
As part of a project on prototyping immersive visualizations of artificial neural network (ANN) 3D node-link diagrams \citep{Behery2023}, the need for a performant and flexible line rendering solution in UE became apparent.
The node-link diagrams consist of thousands to millions of world-space lines, which are potentially updated dynamically.
Unfortunately, UE does not provide optimized line rendering functionality, which is uncommon in entertainment contexts.
As efficient line rendering is a recurring requirement in immersive visualization, we decided to implement the feature as a general plugin that can be reused in other scenarios.

While UE provides a way to render larger line batches, the documentation explicitly states that non-debug use is not performant \citep{linebatcher}. 
In the first implementation, we na\"ively tried to use instanced static meshes to render camera-aligned cylinders and quads.
UE 4.23 did not yet support practical ways to set specific data (e.g. color) per instance, and additionally had trouble achieving acceptable framerates for dynamically updated lines numbering more than a few thousand.
However, in UE 4.25 and later, custom per-instance float parameters could be set and read in the respective material, making our workaround obsolete.
Avoiding slow CPU transform updates, we used a 2D texture to store line segment positions, passing this texture to the shader of the instanced mesh renderer.
We used the new custom float functionality to pass an index into the position texture, width, and color values to each instance, as those remained largely unchanged, and no expensive update was required.
With the position texture and respective index, the vertex shader sets the segment position, scale, and orientation. 
To efficiently update the position texture, line segments are stored in a linear array on the CPU.
Any changes to segment positions are applied to the linear array, of which the relevant parts are then uploaded to the GPU. 
In desktop settings, this allowed us to scale up to millions of lines while still supporting hundreds of thousands of dynamically updated lines in high-resolution HMD settings.
The actual performance, however, is extremely use-case dependent as redundant fragment shader invocations due to quad overdraw can occur and are difficult to work around.
\begin{figure}[ht!]
    \centering

    \includegraphics[width=\columnwidth]{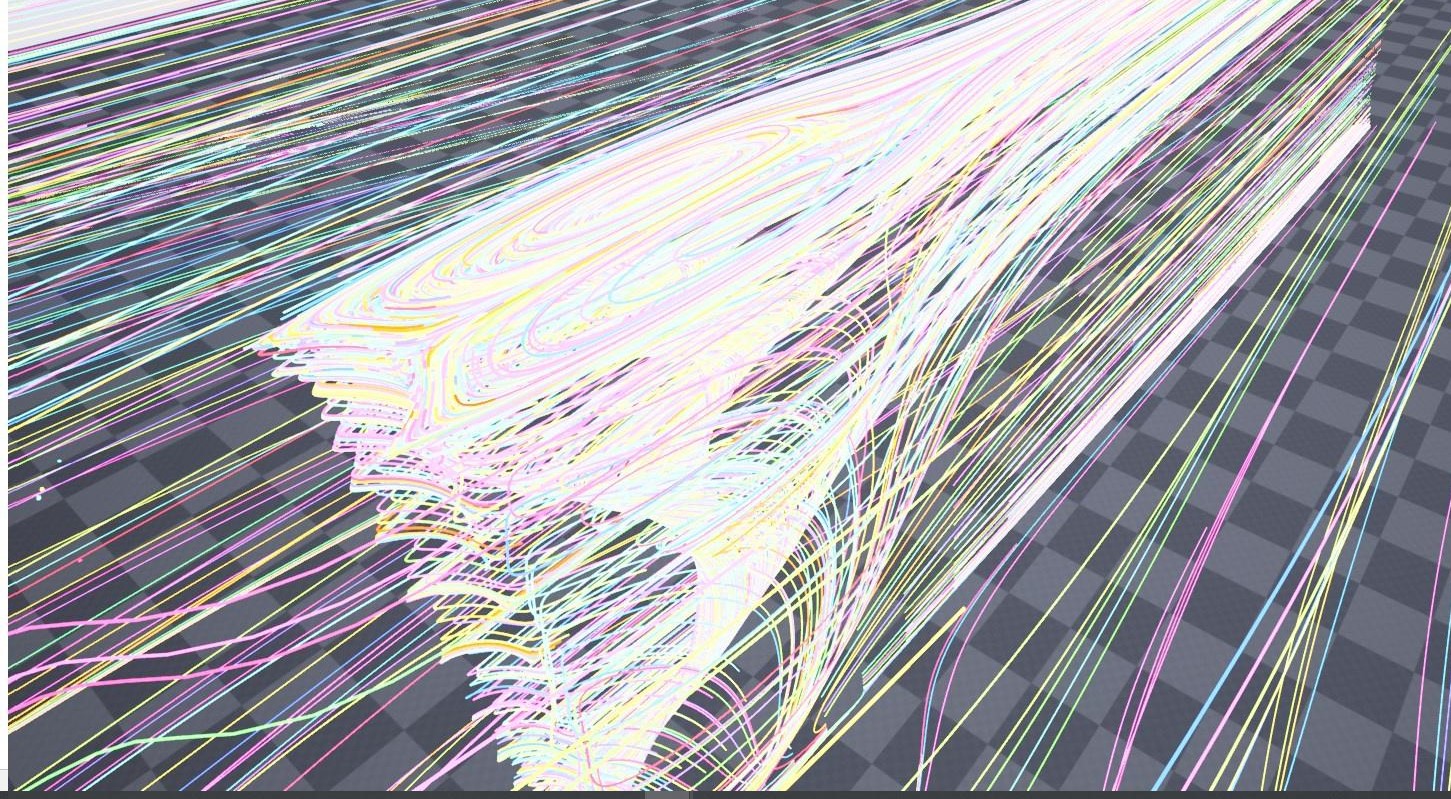}
    \caption{Line rendering of a set of precomputed streamlines using our line rendering plugin.}
    \label{fig:lines}
\end{figure}
Figure \ref{fig:lines} shows a prototype of our line rendering plugin rendering a set of precomputed streamlines.

\begin{figure}[ht!]
    \centering

    \includegraphics[width=\columnwidth]{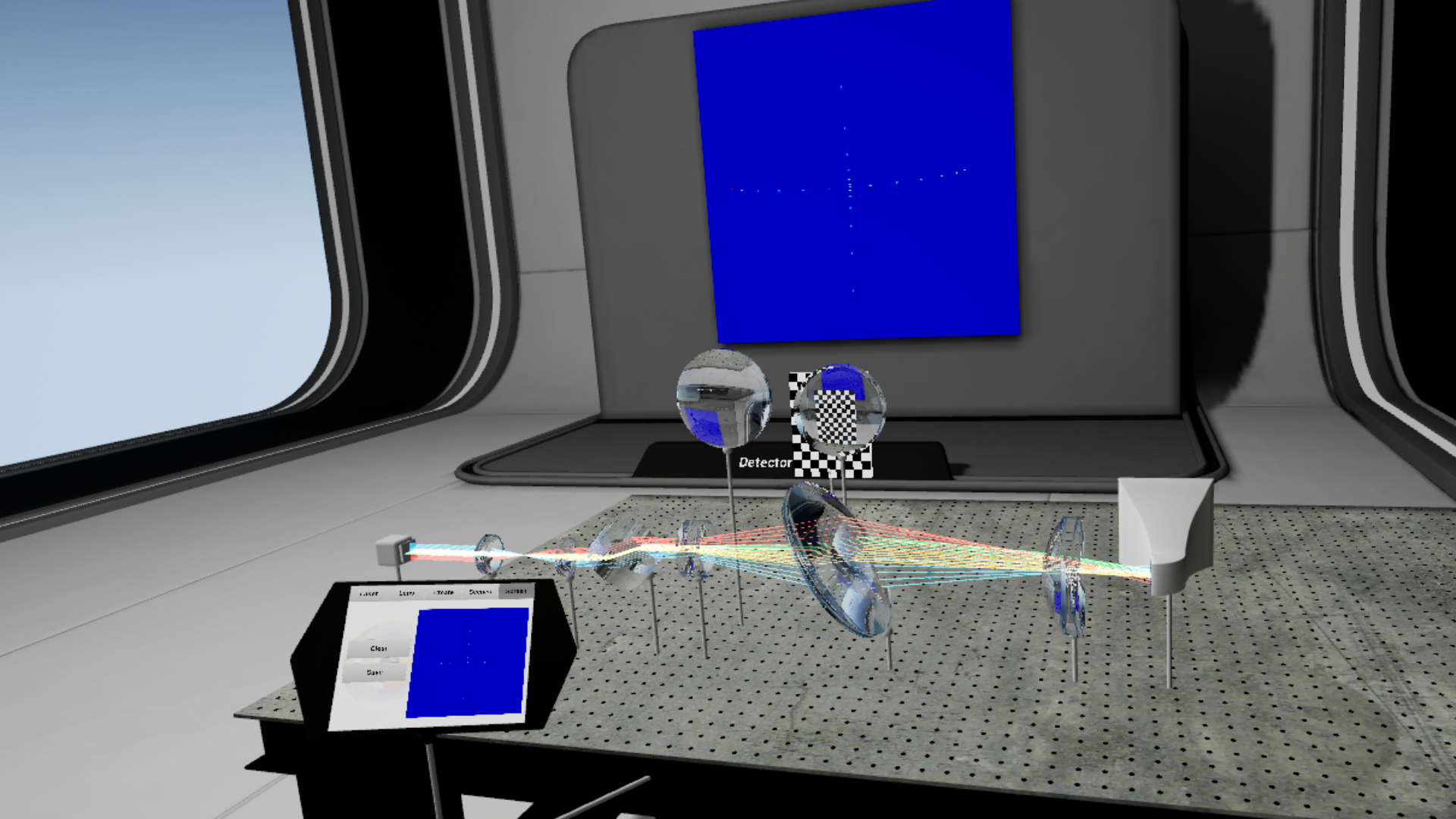}
    \caption{Runtime view of the Virtual Optical Bench, showing an illustrative layout of various lenses. The laser simulation through them and the main viewport rendering pass can be seen. A light intensity detector is projected on the wall and collects the hits of the laser rays.}
    \label{fig:optical_bench_full}
\end{figure}
Besides extending engine functionality, some use cases require to include complete third-party packages and their embedding in UE's provided functionality. 
We highlight one use case that shows the extendable performance of UE by including NVIDIA OptiX in the rendering loop. 

\paragraph{3. Integration with NVIDIA OptiX}
One of our earliest experiences with UE was the development of a Virtual Optical Bench \citep{bellgardt2023optical}.
Originating from a collaboration with the Chair for Technology of Optical Systems at RWTH Aachen, the Virtual Optical Bench is an immersive VR application that is used in a lab course for teaching optical layouts to students.
Experiments on a real optical bench often involve expensive specialized equipment and time-consuming preparations, and can be quite dangerous when not supervised carefully.
Therefore, a simulation was developed and embedded in an immersive virtual environment that visualized the optical effects of spherical lenses and laser ray propagation through them. 
The work used \textit{NVIDIA OptiX} for real-time ray tracing of implicitly defined spherical lenses embedded in a custom framework \citep{pape2021optical}.
NVIDIA OptiX provides a hardware-accelerated raytracer that supports the definition of implicit geometries and achieves acceptable performance in this scenario.
Rendering consists of three parts: First, a regular rasterization pass exists that renders the scene geometry, such as the environment, table, and user interfaces.
Second, one OptiX raytracing pass per eye over the full HMD viewport is performed, which only traces the lenses and various optical targets while using a pre-rendered cubemap to approximate the rasterized environment.
The third pass consists of additional OptiX ray traces to simulate a laser with customizable properties, returning a buffer of lines that are then rasterized. 
Depth results of all three passes are used to achieve correct blending.
The initial implementation was developed in a custom framework, and an expert evaluation yielded initial feedback.
Due to our transition to UE and difficulties deploying the prototype to new hardware for the designated lab course, the application was ported to UE while integrating the collected feedback.
The resulting port can be seen in Figure \ref{fig:optical_bench_full}, showing a virtual lab bench with several lenses on the test bench.
Virtual laser beams are emitted in a cross-pattern, and their propagation is simulated physically accurate through the lenses. 
The beams are steered to a virtual light intensity detector that can display the intensity and distribution of laser beams on its receiver surface, mirrored to the wall for easier visibility. 
\begin{figure}[ht!]
\centering
\includegraphics[width=\linewidth]{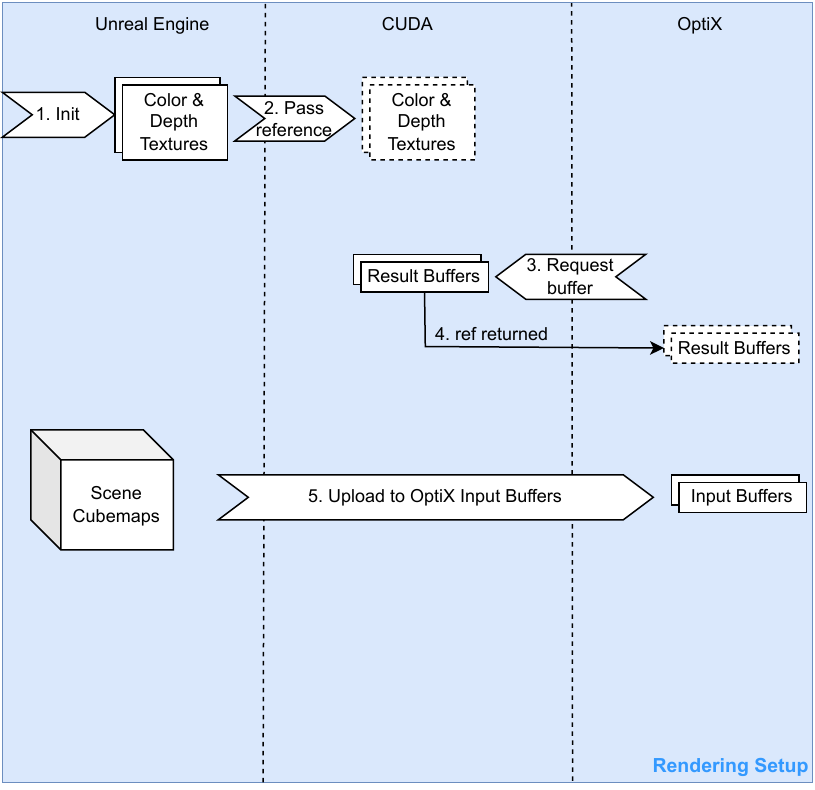}
\hfill
\includegraphics[width=\linewidth]{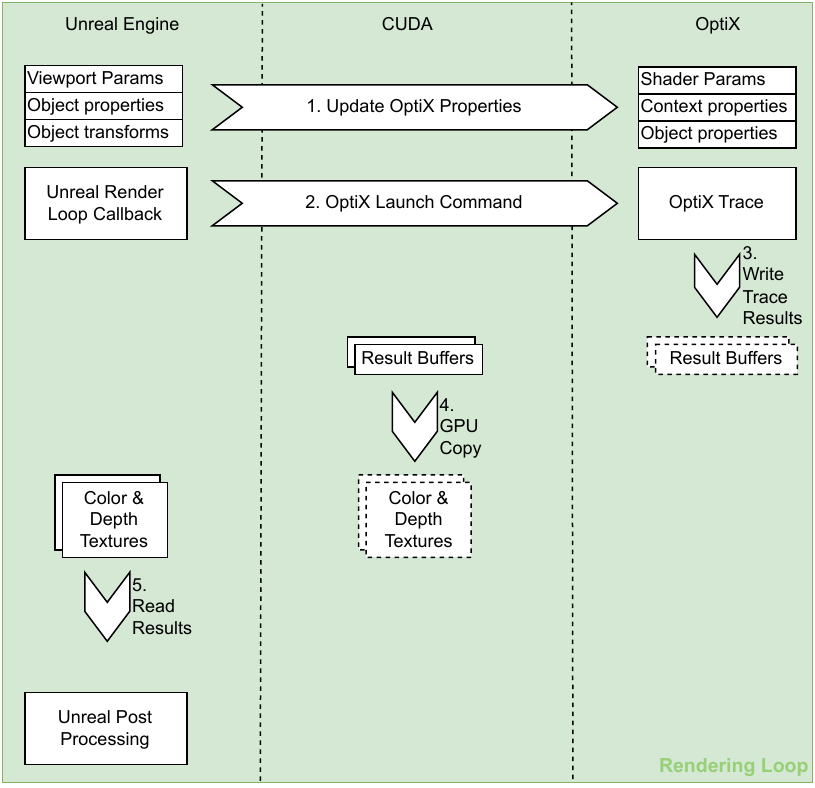}
\caption{Left: Simplified setup of the OptiX and CUDA connection to UE. CUDA is used for interoperability between UE textures and sets up GPU memory buffers for OptiX. Cubemaps are directly uploaded to OptiX input buffers. Right: Simplified rendering loop of the optical bench plugin. Runtime changes to lenses, targets, and simulation properties are passed from UE to the OptiX context. An OptiX trace is then executed from the UE render loop, and the results are copied back into native UE textures via CUDA. These textures are then blended in a regular UE post-processing step.}
\label{fig:optix_pass}
\end{figure}

The integration of OptiX into UE posed three main challenges:
First, we needed to include OptiX and CUDA as third-party plugins into UE and synchronize the transformation matrices of the lenses with geometrical representations in UE.
This allowed interaction with the lenses directly, as we wanted to keep the development of interaction methods contained to UE.
OptiX objects were wrapped with engine primitives, allowing the synchronization of their transform and property changes to the OptiX context.
Since OptiX has to render the result of the viewing rays in the scene context, OptiX needed to be supplied with the rendering parameters from UE and a scene description.
To approximate the results of rays missing the OptiX geometries, one cubemap per lens was captured in UE at the position of the lens and uploaded into an OptiX buffer, yielding a comparable result when no OptiX object was hit.
Because there exists no native interoperability between UE and OptiX, we used CUDA to directly access the buffer memory of OptiX and copy the color and depth render pass results into the native DirectX11 textures exposed by UE's rendering interface for both viewing rays and laser beams.
The textures were then used in a post-processing material to blend the results with UE's basepass framebuffer, which included the typical rasterized scene. 
The simplified setup and loop of the custom viewport render pass are schematically shown in Figure \ref{fig:optix_pass}.
The ray behavior was copied from the texture, and consequently, rays were rendered natively in UE.
As an overarching difficulty, the visualization needed to run at acceptable framerates for HMDs.
Unreal Insights allowed the incorporation of custom scoped performance traces into our rendering code, highlighting potential bottlenecks.
The access to internal engine features was key for developing an approach that combines CUDA and UE's rendering interface, avoiding expensive memory and texture copies from the GPU device to the host CPU. 
Furthermore, separating UE's scene geometry and rendering from the raytraced OptiX geometry yielded the best possible acceleration structures for the respective passes.

In summary, developing this extension showed us how powerful a deep and complex third-party software integration into UE can be. 
It allowed us to benefit from all of UE's large feature set regarding basic scene rendering, interaction development, and platform flexibility while maintaining the respective accessibility and performance of the scene rendering.
Meanwhile, we also reaped the benefits of the hardware-accelerated raytracing from NVIDIA OptiX, which would otherwise have been impossible in UE then.
In that regard, the high initial time investment of investigating where and how we could couple the two systems paid off, giving us a powerful application that is used in actual laboratory courses.

\subsection{Impact on Daily Work}
So far, we only presented our utilization of UE in particular technical use cases.
However, the wide adoption and accessibility majorly impacted our teaching activities and immersive visualization services. 
At the same time, the provided tooling benefited all development areas and turned out to be one of the main bonus points of UE.

\paragraph{1. General Tooling}
As discussed in the requirement section, tooling is a property commonly overlooked in research environments.
While it is often not a hard requirement, it can help developers accelerate their development and testing process across the board. 
The flexibility of UE's build system helped us set up our CI pipeline based on the community work by Adam Rehn \citep{unrealdocker}, showing the benefits of using widely adopted COTS software also with regard to tooling. 
Our pipeline provides building via continuous integration on our centralized build servers.
This allows us to free up resources on the developer machines and automate the deployment of applications to our CAVE.
Tooling such as UE's Derived Data Cache server allows us to set up a lab-wide build cache solution that further speeds up development.
UE additionally provides a distributed light-building system that accelerates the offline process of light baking by distributing it on multiple machines. 

Going beyond simplifying the build process, UE's Multi-User Editing allows collaboration on a project simultaneously, avoiding time-consuming merges.
The Multi-User Editing is also integrated into nDisplay, making it possible to develop and edit scenes running live on our CAVE environment, cutting down on iteration times and directly giving us an impression of the scene in CAVE-based virtual environments. 
For setting up and launching nDisplay and multi-user editing applications, UE's Switchboard tool can be used. 
The extensibility and adaptability of UE also apply to the provided tooling.
To use Switchboard on our Linux environment, we modified the tool to pass specific flags to application instances running on our nodes.

\begin{figure*}[ht!]
\centering

\includegraphics[width=0.91\linewidth]{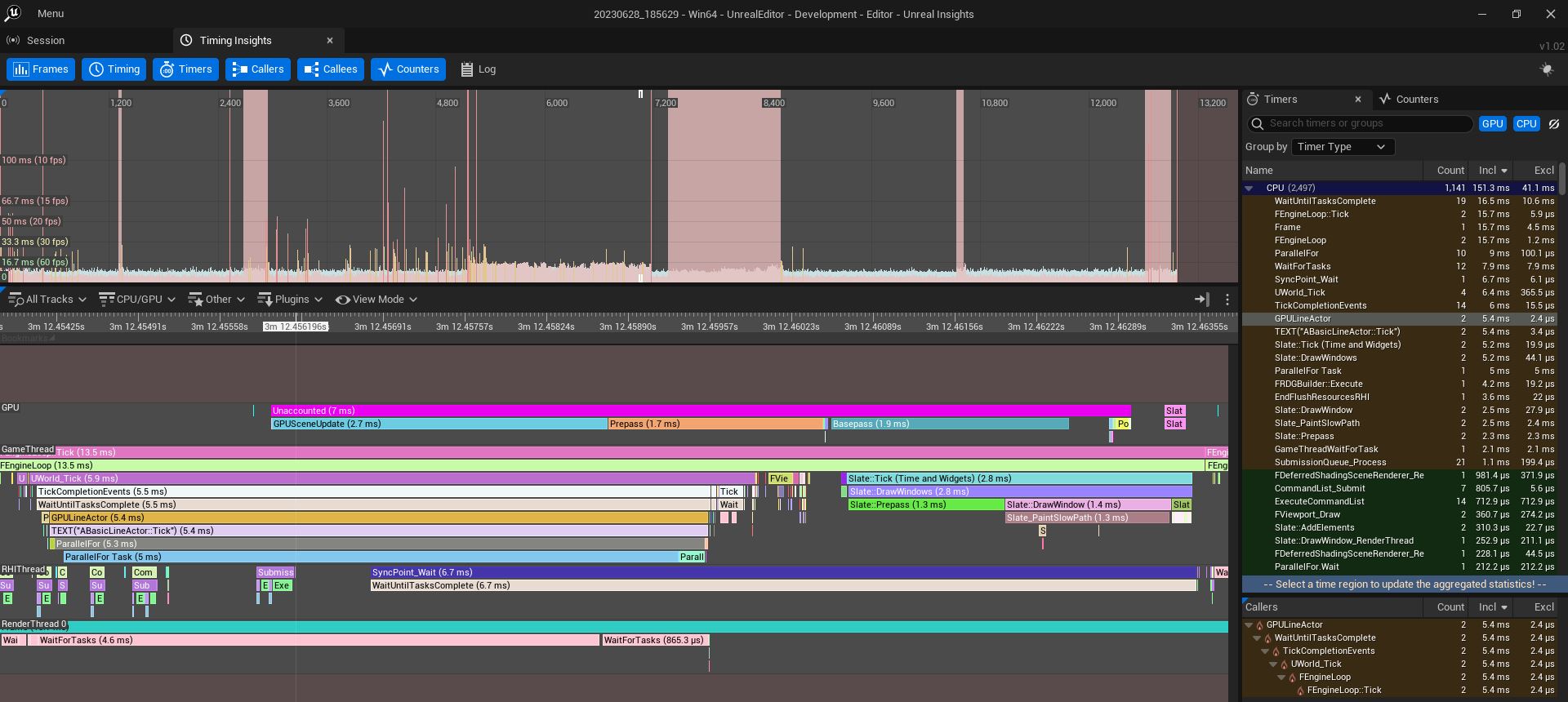}

\caption{Unreal Insights shows a custom trace when debugging performance of our line rendering plugin. }
\label{fig:insights}
\end{figure*}

Unreal Insights, a performance measuring tool that can get performance information about almost every aspect of the engine and user code, also simplified performance tweaking in our projects, such as our line rendering plugin, as seen in Figure \ref{fig:insights}. 
It was a key component in helping us track down an issue in the nDisplay threading code and keeping our targeted framerates for immersive VR applications. 
If low-level GPU debugging is required, the RenderDoc \citep{renderdoc} plugin can be used to inspect UE's rendering passes, which is very helpful when extending the native rendering or debugging issues.
Developing such tools with the same quality, integration, and features for our own custom framework with the limited time and developer resources would be impossible.
 
\paragraph{2. Teaching Activities}
As our institute offers bachelor and master theses as well as a semester-long practical course, students need to familiarize themselves quickly with the game engine to be productive. 
The wide adoption of UE throughout the industry and in research has led to many free online tutorials in video and text form, with over 2500 articles being available on the official Unreal developer community platform \citep{epiccommunity}.
This starkly contrasted our challenges in setting up students with previous frameworks.
Students often needed more knowledge to build ViSTA, requiring time-consuming supervision and onboarding processes.
With UE, they can instead find help and solutions to common problems in the community and documentation. 
Additionally, UE offers the Blueprint visual programming language.
Compared to the pure C++ interface of ViSTA, this lowered the barrier of entry further, as students often require fast prototyping to investigate potential solutions early in their theses.
By offering both Blueprints and the possibility to write computationally intensive work in C++ code, UE achieves the required accessibility for our students while still fulfilling the performance criteria in case specialized approaches are required.
Finally, the flexibility of UE regarding HMDs provided students with the ability to use their personal headsets at home, as platform support was already available in UE.

\paragraph{3. Immersive Visualization Services}
Besides teaching activities and research, UE is used in our lab to provide immersive visualization services to other institutes and third parties.
Due to the regular short time academic and non-academic staff is employed, the general accessibility of UE also helps new staff members get up to speed, leading to lower onboarding times and higher productivity. 
The wide adoption of UE often leads to cooperation with partners already aware of UE, making communication easier by building upon a common ground.
Sizeable portions of requests are 3D data visualization tasks where data provided by cooperation partners often come in formats common in the respective domains of the partners.
UE supports a plethora of different 3D mesh data natively or via plugins such as the UE Datasmith tool, saving ample development time for custom data importer software.
While the meshes are often not optimized and require further manual refinement, they provide a good starting point, which would otherwise not exist.
Manual refinement is directly supported by UE features that allow optimization in a comprehensive editor, thus reducing the development time needed to provide services.
Switching from a custom framework to UE thus freed tremendous resources from our service team regarding the maintenance and development of the custom framework itself.

\section{Discussion}
After outlining the impact of UE in several example projects, we discuss trade-offs and considerations for using game engines.
Over time, the large feature set provided by the engine evolved as the most significant advantage and argument for applying game engines in immersive visualization research.
It is of utmost necessity to promptly adapt to different use cases and domains while staying current with cutting-edge research and technology.
UE provides an excellent foundation to quickly build upon and deliver immersive solutions for various domains, from scientific visualization to visual analytics, drastically reducing development times.
It further allows efficient \textit{What You See Is What You Get} (WYSIWYG) editing for UI development and scene authoring.
The editor provides methods to easily author scenes that prove helpful for projects such as the aerosol visualization.
Multi-User-Editing allows live changes to be directly projected into the CAVE, yielding immediate visual feedback on the target platform.
The UI editor allows the creation of complex UIs intuitively by combining extendable UI primitives.

However, the caveat is that much functionality is mostly tailored toward game development, and only a few ready-made solutions for typical immersive visualization applications are applicable out of the box.
For example, UE does not provide ready-to-use functionality for selecting and manipulating objects or navigation in virtual environments.
Our lab developed an open-source VR Toolkit for UE~\citep{rwthtoolkit} to address this issue, which includes standard 3D interaction and navigation techniques customized for research applications and purposes.
However, the additional maintenance and development costs must be considered when maintaining such a toolkit, as changes in the UE API between versions regularly lead to necessary changes for compatibility. 
Maintaining support for multiple engine versions simultaneously has shown to be challenging, as old and new features must be tested and kept compatible with all supported versions.
The only alternative is to add new features only for newer engine versions, neglecting backward compatibility, which leads to divergent feature sets regarding engine versions.

As described in the previous section, some features of typical immersive visualization applications require access to core functionality, which is sometimes inaccessible in applications by default.
Due to the access-to-code policy by UE, it is possible and encouraged to adapt the engine code to individual purposes, which also allows the exposition of the required functionalities if needed.
Nevertheless, changes to the engine must be carefully considered for several reasons.
First, due to the large feature sets and provided tooling, the UE codebase is extensive and intimidating, making it challenging to understand the general structure.
This especially applies to core functionality written in optimized and abstracted code, such that even small changes can be time-consuming.
Second, modifying the engine makes version upgrades substantially more complex. 
New engine releases require a manual merge of the changes, fixing compile issues, and extensive testing to ensure no bugs were introduced.
Lab infrastructure becomes more complex as custom source builds must be deployed on all machines.
Third, changes in the engine core have a heightened potential to introduce performance regressions.
Finally, a customized engine makes collaboration with third parties more difficult, as the customized engine version must be used, thus reducing the advantages of large adoption.
Considering this, we believe that engine source modifications should be kept to a minimum.
They should be limited to specific controlled environments such as the CAVE if absolutely necessary.

Further, we strive to contribute any changes upstream into the official UE codebase.
This allows the community to benefit from the contributions and enables us to fall back to official releases once merged.
One of these examples was the necessary modification for compatibility with our CAVE system.
The necessary changes to support Linux and later Vulkan for nDisplay were contributed as pull requests upstream. 
Linux support of nDisplay was officially added to the UE starting from version 4.27, allowing us to return to the unmodified official release.
We experienced that Epic Games values merge requests for UE provided by the community, as several of our merge requests were accepted.
Epic Games' overall development process can, however, appear opaque.
It is not always clear which new features are worked on and will appear in the subsequent releases, especially when niche, performance-critical aspects are considered.
The official repository provides the only real insight where current development efforts can be tracked.
This can lead to situations where official releases render in-house solutions obsolete.

Epic Games supports community development efforts by offering 'Epic MegaGrants' and providing financial support for community developers.
With the large community of developers, labs can benefit from and contribute to the engine's ecosystem.
Community-to-community help on Discord servers and in forums, free and open-source plugins, and the availability of online resources are beneficial for labs.
Exchanging experiences in the community and other labs is especially important for non-standard use cases, such as CAVEs.
While the various forms of community-to-community help are welcome, they also contain some inherent risks.
We observed, for example, that existing community resources are often outdated, incomplete, or of low quality.
This applies to both source code for community plugins and online learning resources, such as tutorials.
As described in our case study \citep{kruger2023case}, community tutorials are often tailored towards proof of concept implementations for use in smaller scales in games and rarely describe the use of advanced features.
However, these advanced features are often necessary to develop performant applications, as seen in line rendering, aerosol/combustion visualization, or the linked-view application for neuronal simulation.

The broad adoption of the engine also leads to unclear best practices, as many alternative approaches are described when comparing official and community resources.
While the official documentation is often lacking details and complex implementation guidance, other accessibility aspects of the engine can be extremely beneficial:
The availability of learning resources and features like the visual programming provided by Blueprints further reduce the entry barrier and allows for rapid prototyping.
Due to the multitude of projects developed in an immersive visualization lab, the ability to quickly provide an easily iterable prototype is crucial.
It allows us to quickly present and discuss first results to cooperation partners, test new research approaches, and subsequently iterate over the solution.

However, developers must keep performance in mind and use the visual programming features sparingly, as these can have severe performance implications, which we could observe in several of the use cases listed above.
In our experience, the best results are obtained using a good balance between C++ implementation and Blueprints.
Computation and repetition-heavy code, e.g., for-loops, benefit immensely from being implemented in C++.
As UE's architecture allows to easily expose C++ implementation to Blueprints, performance-critical or computation-heavy code can be easily implemented in C++ while initiating the call, and further processing might be achieved in Blueprints.
Besides custom C++ implementations, UE provided functionality for Blueprint nativization until version 5.
This means that blueprint code is mostly compiled into C++ code with a significantly smaller overhead.
Beginning with version 5.1, UE supported the porting of Blueprint code to C++ code with new functionality in the engine, making the transition from Blueprint to C++ implementations easier.

The provided flexibility that balances rapid prototyping and performance is especially noticeable in teaching and services, as previously described.
Besides Blueprints, UE provides several more features that increase development speed and make development easier.
Shaders can be programmed in UE's material editor, which provides a visual programming interface to create materials and offers a live preview of the current result.
This gives programmers a comfortable way to evaluate the current shader while developing.
In some cases, however, we experienced that the material editor can make shader development more opaque as it is hard to judge what the resulting shader will look like once compiled by UE.
This is especially apparent when the shader has to perform tasks beyond standard shading, e.g., the calculations and optimizations needed for the line rendering.

Revisiting and evaluating the six requirements proposed in the requirements section, we come to the following conclusions to date:
\begin{description}[wide, labelwidth=!, labelindent=0pt]
    \item \textbf{Requirement 1 - Wide Adoption} The assumptions we put into the advantages a wide adoption brings have been shown to be mostly correct. 
    New work staff and students often mention already in their applications that they are familiar with UE, cooperation partners are familiar with the software, and we both use and provide functionality via community plugins.
    We, therefore, evaluate this requirement as \textit{fully met}.

   \item \textbf{Requirement 2 - Large Feature Set} As described in the use cases, we use various native engine features to develop applications. 
   This allows us to develop new applications quickly and easily, and the tooling provides significant support while developing. 
   However, features are often focused more towards game development, such that functionality needed for data visualization, especially immersive visualization, is often not provided out of the box.
   The general feature richness allows, however, the development of missing functionality rather easily.
   Unreal Engine undoubtedly has a large feature set; however, as typical visualization functionality is mostly not provided by the engine, we evaluate the large feature set as \textit{mostly met} for the purpose of developing immersive visualization applications.
   
    \item \textbf{Requirement 3 - Performance} In our experience, UE can provide fantastic performance due to its highly efficient rendering core and state-of-the-art rendering techniques such as Nanite.
    We acknowledge and stress, however, that best practices, experience, and due diligence are necessary to reach maximum performance. 
    Due to this, we evaluate this requirement as \textit{met under conditions}.
    \item \textbf{Requirement 4 - Accessibility}
    We noticed a decrease in the need for help for students and new colleagues after switching to UE.
    We attribute this mainly to available resources and features provided by the engine that help new developers.
    As discussed in the previous paragraph, the quality of documentation, both official and community-provided, can vary and lead to suboptimal decisions and implementations, as shown in~\citet{kruger2023case}.
    We, therefore, deem this requirement as \textit{met under conditions}, w.r.t. immersive visualization, as guidance by experienced colleagues is still necessary.
    \item \textbf{Requirement 5 - Extensibility/Adaptability}
    As UE allows not only the implementation of features in user code but also provides the ability to change the engine's source code directly, we were able to tailor UE to our needs.
    This was shown in various use cases, most notably in the \textit{Adaptation for AixCAVE}, which required deep extensions and changes inside of the engine.
    Therefore, the requirement is \textit{fully met} by UE.
    
    \item \textbf{Requirement 6 - Flexibility} As described in the \textit{Unreal Engine In The Lab} section, we run applications on a multitude of hardware paired with different operating systems.
    The development experience is overwhelmingly uniform, with most changes occurring in the packaging stage when applications are built for the target system.
    We, therefore, evaluate this requirement as \textit{fully met}.
\end{description}

The following table provides a comprehensive overview of the fulfillment of the requirements:

\noindent
\begin{table}[h!]
\begin{tabularx}{\columnwidth}{@{}Xcccccc@{}}
            & Req. 1 & Req. 2 & Req. 3 & Req. 4 & Req. 5 & Req. 6 \\ \hline
Fulfillment &    FM    &     MM   &   MUC      &   MUC     &  FM  &    FM \\ \bottomrule
\multicolumn{7}{c}{\scriptsize FM: Fully Met $\bullet$ MM: Mostly Met $\bullet$ MUC: Met Under Conditions} \\
\end{tabularx}
\caption{Tabular overview of the degree of fulfillment UE achieves for each requirement.}
\end{table}

\section{Conclusion and Future Work}
In this work, we shared our experiences and considerations about using Unreal Engine as a primary means of developing and researching immersive visualization applications and techniques at our immersive visualization lab at RWTH Aachen University.
As game engines developed to be more versatile, widespread, and performance-driven, they also became a suitable consideration for developing complex immersive visualization applications. 
First, we briefly covered how developing custom frameworks for immersive visualization for several years led to our decision to use UE as the primary software base.
We then presented and justified six qualitative requirements for game engines developed from our experiences.
These are wide adoption, large feature set, performance, accessibility, extensibility/adaptability, and flexibility. 
Using examples from our lab, we showed that UE can fulfill these requirements and is thus a suitable choice for immersive visualization labs.
However, we were also able to show limitations and challenges that became evident and provided insights into how we were able to address them.
After discussing and weighing these aspects, UE provides a solid base for developing immersive applications.
It further provides a lowered entry barrier, which has shown to be extremely useful for new and experienced developers.
While we are still investigating and evaluating specific aspects of the engine, especially concerning standard visualization data formats, algorithms, and methods, our experience is mostly positive.
Though some aspects can be challenging when working with UE, the benefits of the large community, performance, accessibility, and flexibility clearly outweigh the challenges.
Therefore, we will use and build upon UE as the primary game engine for the foreseeable future while continuously evaluating our experiences with regard to immersive visualizations.

There are, however, still some open aspects that we would like to investigate in the future.
First, detailed insights into the complexity, difficulty, and benefits of implementing core rendering extensions would be helpful.
This is especially interesting for rendering-heavy problems such as line rendering or advanced rendering techniques not covered by the existing implementation.
Investigations into the capabilities of running code that performs typical visualization tasks such as particle advection would be extremely viable.
Using UE's compute shader support for hardware-accelerated run-time generation of meshes based on, e.g., distance fields and iso-values is of high interest, as it would empower the engine to compute results and simulations based on interactive parameters dynamically.
During the presentation of the use cases, the lack of standard tooling for standard formats in visualization was already mentioned.
While we discussed our workflows to circumvent these restrictions, we want to investigate further how such formats can be used directly with UE.
Similarly, we are interested in combining UE with the established method of in-situ/in-transit approaches to provide immersive in-situ/in-transit experiences.
The direct integration of the AixCAVE into the RWTH compute cluster allows us to nicely combine our lab's existing ease-of-use focused research on in-situ/in-transit \citep{kruger2023insite,kruger2023insiteposter} with UE.
A quantitative analysis of the performance limitations and comparison to custom approaches would provide a basis for a more informed decision-making process.
Quantified results and analysis would facilitate judgment on which techniques can be developed with Unreal Engine as a base for immersive visualization applications and identify use cases that cannot be covered with Unreal Engine. 
Finally, a thorough evaluation of other COTS software besides Unreal, such as the similarly popular Unity engine, regarding our presented requirements, could provide a better overview of possible alternatives and comparisons.
\bibliographystyle{apacite}

\bibliography{main}
\end{document}